\documentclass[12pt,preprint]{aastex}
\usepackage{amsmath,amssymb}

\begin{document}

\title{Modeling sunspot and starspot decay by turbulent erosion}

\author{Yuri E. Litvinenko} \affil{Department of Mathematics, University of
Waikato, P. B. 3105, Hamilton, New Zealand}

\author{M. S. Wheatland} \affil{Sydney Institute for Astronomy, School of Physics,
The University of Sydney, NSW 2006, Australia}

\begin{abstract}
Disintegration of sunspots (and starspots) by fluxtube erosion, 
originally proposed by Simon \& Leighton, is considered.
A moving boundary problem is formulated for a nonlinear 
diffusion equation that describes the sunspot magnetic field profile. 
Explicit expressions for the sunspot decay rate and lifetime 
by turbulent erosion are derived analytically and verified numerically. 
A parabolic decay law for the sunspot area is obtained. 
For moderate sunspot magnetic field strengths, the predicted decay rate 
agrees with the results obtained by Petrovay \& Moreno-Insertis. 
The new analytical and numerical solutions significantly improve 
the quantitative description of sunspot and starspot decay by turbulent erosion. 
\end{abstract} 


\section{Introduction}

Bumba (1963) investigated how the areas of large,
slowly decaying sunspots decrease with time. 
His data analysis suggested that the sunspot area $A$ decreases linearly 
with time $t$: 
\begin{equation}
  A(t) = A_0 - \dot{A} t , 
\end{equation}
where the decay rate $\dot{A}$ is a constant. 
The result is consistent with the Gnevyshev--Waldmeier relation 
$T \sim A_0$, where $T$ is the sunspot lifetime 
and $A_0$ is its initial area 
(see, e.g., Petrovay \& van Driel-Gesztelyi (1997) for a review). 
Following Bumba (1963), sunspot observations were usually interpreted 
in terms of the linear decay law for the sunspot area 
(e.g., Robinson \& Boice 1982). 
Yet it is difficult to distinguish linear and nonlinear decays 
observationally, and observations have also been 
interpreted using a parabolic decay law, 
with $A(t)$ a decreasing quadratic function of time 
(Moreno-Insertis \& V\'azquez 1988; 
Mart\'inez Pillet et al. 1993). 

On the theoretical side, Meyer et al.\ (1974) argued that 
the linear decay law is a consequence of turbulent diffusion of 
the magnetic field across the whole area of a sunspot and 
expressed the constant decay rate in terms of a constant uniform 
diffusivity (see also Krause \& R\"udiger 1975). 

Simon \& Leighton (1964) inferred from observations that 
the gradual disintegration of sunspots is due to ``erosion'' 
of the penumbral boundaries by supergranular flows, 
which occurs when bits of magnetic field are sliced away 
from the edges of the sunspot and swept to the supergranular 
cell boundaries. In contrast to the model of Meyer et al.\ (1974), 
such erosion can occur if the turbulent diffusivity associated with 
the flows is suppressed within the spot (Petrovay \& Moreno-Insertis 1997). 
Alternative theoretical approaches were reviewed by Solanki (2003). 

Petrovay \& Moreno-Insertis (1997) developed the turbulent 
erosion model mathematically, taking into account the dependence 
of the turbulent diffusivity on the magnetic field strength. 
The diffusivity rapidly decreases if the magnetic field exceeds 
an energy equipartition value (Kitchatinov et al.\ 1994). 
As a result, a current sheet is formed around the spot. 
The model leads to the parabolic decay law, specified by 
a constant inward speed $w$ of the current sheet, viz., 
\begin{equation} 
  A(t) = \pi (r_0 - wt)^2 
\end{equation} 
for a circular flux tube (sunspot) of an initial 
area $A_0 = \pi r_0^2$. Moreover, the model yields $w \sim 1/r_0$, 
and so it agrees with the Gnevyshev--Waldmeier relation. 
Petrovay \& Moreno-Insertis (1997) concluded that solar observations 
are consistent with turbulent erosion based on a granule-size 
diffusion length. Petrovay \& van Driel-Gesztelyi (1997) 
presented observational evidence in favor of the parabolic decay rate, 
predicted by the turbulent erosion model, although an independent 
magnetohydrodynamic simulation suggested that 
the sunspot decay law is almost linear (R\"udiger \& Kitchatinov 2000). 
Petrovay et al.\ (1999) also explored the effect 
of a preexisting ``plage'' field on the decay rate, whereas 
Chatterjee et al. (2006) applied the model to the development of twist 
in a flux tube rising through the solar convection zone. 

The analytical results of the turbulent erosion model have been recently 
used to complement numerical simulations of sunspot formation and decay 
(e.g., Rempel \& Cheung 2014). 
The model has also been applied to starspots, 
with a goal of using the starspot decay data 
to place constraints on the magnetic diffusivity, 
which may be useful for dynamo models 
(e.g., Strassmeier 2009; Bradshaw \& Hartigan 2014). 

It is worthwhile to revisit the turbulent erosion model of sunspot decay. 
The original calculation of Petrovay \& Moreno-Insertis (1997) 
was guided by numerical results and one-dimensional analytical solutions. 
A dimensional argument was used to estimate 
the magnetic field gradient at the sunspot edge: 
\begin{equation} 
 \frac{\partial B}{\partial r} \sim - \frac{B_e}{r_0} , 
\label{eq-PM-scaling}
\end{equation} 
where $B_e$ is the magnetic field value above which 
the turbulent diffusivity is assumed to be suppressed 
(see equations (11) through (16) in Petrovay \& Moreno-Insertis 1997). 
In addition, their numerical estimate for the sunspot lifetime 
appears to be based on an estimate of the current sheet speed $w$ 
rather than on direct computation. 

A rigorous derivation of the sunspot decay law is necessary 
if the theory is to be used to develop reliable predictive tools. 
Explicit analytical predictions of the turbulent erosion model 
could complement more detailed numerical (e.g., Hurlburt \& DeRosa 2008) 
and empirical (Gafeira et al. 2014) models of sunspot decay. 
Hence our aim is to put the turbulent erosion model on a firmer footing. 
We do this by formulating a moving boundary problem 
(Carslaw \& Jaeger 1959; Crank 1984) for the model 
and solving it to derive a prediction for the sunspot decay law. 
In the remainder of the paper, we present the new analytical (Section 2) 
and numerical (Section 3) results and their discussion (Section 4).

\section{Formulation of the problem and analytical results}
\label{sec:analytical_model}

In order to model the turbulent erosion of a sunspot, 
we follow Petrovay \& Moreno-Insertis (1997) and consider 
the evolution of a cylindrically symmetric magnetic flux tube. 
The magnetic field ${\bf B} = B(r, t) \hat{\bf z}$ 
is described by the diffusion equation 
\begin{equation}
  \frac{\partial B}{\partial t} = 
  \frac{1}{r} \frac{\partial}{\partial r} 
  \left( r D \frac{\partial B}{\partial r} \right) . 
\label{eq-axisymm-diff}
\end{equation}
Here $t$ is time, and $r$ is the distance from the $z$-axis. 

The turbulent diffusivity $D = D(B)$ is strongly suppressed when 
the magnetic field exceeds an energy equipartition value 
$B_e = \sqrt{4 \pi \rho} u$ where $\rho$ is the mass density and 
$u$ is a characteristic turbulent speed (e.g., Kitchatinov et al.\ 1994). 
For instance, taking a photospheric value of 
$\rho \approx 2 \times 10^{-7}$ g cm$^{-3}$ and a granular value of 
$u \approx 2 \times 10^5$ cm s$^{-1}$ yields $B_e \approx 400$ G 
(Petrovay \& Moreno-Insertis 1997). 
To simplify the analytical treatment, we assume 
\begin{equation}
  D(B) = D_0 = \mbox{const}, \quad B < B_e 
\label{eq-diffusivity}
\end{equation}
and 
\begin{equation}
  D(B) = 0, \quad B > B_e. 
\end{equation}
The initial value problem is specified by the field profile 
\begin{equation}
  B(r, 0) = B_0 = \mbox{const}, \quad 0 < r < r_0 , 
\end{equation}
and $B(r, 0) = 0$ otherwise (see Tlatov \& Pevtsov (2014) for recent data 
on sunspot magnetic fields). We nondimensionalize the problem 
by measuring the magnetic fields, times, and distances 
in units of $B_e$, $r_0^2/D_0$, and $r_0$, respectively. 

The sunspot size decreases with time 
because the magnetic flux is removed by diffusion. 
The strongly nonlinear dependence of the diffusivity $D$ 
on the magnetic field strength leads to the formation of 
a tangential discontinuity at the edge $r = r_e$ of the flux tube. 
Physically, the magnetic field discontinuity at $r_e (t)$ 
corresponds to a current sheet at the sunspot edge, 
where the magnetic flux removal is made possible 
by a strongly localized electric current. 

It is useful to observe that the problem at hand is 
mathematically similar to a moving boundary problem in the theory of 
heat conduction, and so we can use existing methods of analysis. 
In particular, an analog of the Stefan condition is obtained 
by the integration of the governing diffusion equation 
across the moving boundary $r = r_e(t)$ (Carslaw \& Jaeger 1959). 
Allowing for the tangential discontinuity at $r = r_e(t)$, 
we substitute 
\begin{equation}
  B(r, t) = B_0 + (B - B_0) H[r - r_e(t)] , 
\end{equation}
where $H$ is the Heaviside step function, into equation (\ref{eq-axisymm-diff}) 
and integrate across the discontinuity (from $r_e-0$ to $r_e+0$). 
The result is 
\begin{equation}
  (B_0-1) \frac{d r_e}{dt} = 
  \left. \frac{\partial B}{\partial r} \right|_{r=r_e+0} , 
\label{eq-stefan}
\end{equation}
where we used $B (r_e-0, t)=B_0$, $B (r_e+0, t)=1$, 
and $D(r_e-0) = 0$. 

To find an approximate analytical solution, we use 
the pseudo-steady-state approximation that can be adopted 
when the rate of change $\dot{r}_e$ is small compared with 
a global diffusion rate $\sim 1$, making it possible to neglect 
the term $ \partial B / \partial t $ in Equation (\ref{eq-axisymm-diff}). 
Physically, the magnetic field profile near a moving boundary relaxes 
to a pseudo-steady state on a time scale $\delta t_D \simeq (\delta r)^2 / D$ 
where $\delta r \simeq \dot{r}_e \delta t$ 
is the displacement of the boundary $r_e (t)$ in a time $\delta t$. 
The approximation is valid if the relaxation is sufficiently rapid, 
say if $\delta t_D \ll \delta t$. In our dimensionless variables, 
we have $\delta r \simeq r_e \le 1$ and $D = 1$, and so 
$\delta t_D / \delta t \simeq \dot{r}_e$. If $T$ is 
the sunspot lifetime, we use $\dot{r}_e \simeq T^{-1}$ to infer 
that, as long as $T \gg 1$, the approximation is globally valid 
in the range $1 < t < T-1$. We show below that roughly $T \simeq B_0-1$. 
Consequently, the pseudo-steady-state approximation becomes more accurate 
as $B_0$ increases. Detailed analysis of the accuracy of the approximation 
can be found in standard textbooks on heat conduction 
(e.g., Crank 1984; Hill \& Dewynne 1987). 

Inside the spot, the vanishing diffusivity implies that 
the magnetic field is constant: 
\begin{equation}\label{eq-b-interior}
  B(r<r_e(t), t) = B_0 . 
\end{equation}
Outside the spot, the pseudo-steady-state field satisfies 
\begin{equation}
  \frac{1}{r} \frac{\partial}{\partial r} 
  \left( r \frac{\partial B}{\partial r} \right) 
  = 0 , 
\end{equation} 
and so 
\begin{equation}
  r \frac{\partial B}{\partial r} = \mbox{const} . 
\label{eq-steady}
\end{equation} 
Equation (\ref{eq-diffusivity}) gives the boundary condition 
\begin{equation}
  B(r=r_e(t), t) = 1 , 
\end{equation}
which would be $B = B_e$ in dimensional units. 
The magnetic field diffusion outside the spot causes the field 
to become negligibly small at some $r=r_f(t)$ outside the spot. 
Solutions of the standard diffusion equation in two dimensions 
suggest that $r_f(t) = (2t)^{1/2}$ (Carslaw \& Jaeger 1959). 
Thus we set 
\begin{equation}
  B(r=r_f(t), t) = 0 . 
\end{equation}
The solution of equation (\ref{eq-steady}), satisfying 
the boundary conditions at $r_e$ and $r_f$, is given by 
\begin{equation}\label{eq-b-exterior}
  B(r>r_e(t), t) = \frac{\ln (r^2/2t)}{\ln (r_e^2/2t)} . 
\end{equation}
On substituting this into equation (\ref{eq-stefan}), we get 
\begin{equation}
  (B_0-1) \frac{d r_e^2}{dt} = 
  \frac{4}{\ln (r_e^2/2t)} . 
\label{eq-re-ode}
\end{equation}

We also obtained a similar differential equation for $r_e(t)$ 
using an independent heat-balance approximation (e.g., Crank 1984). 
We do not present the results here: although 
the approach requires longer calculations, it does not appear 
to be more accurate than the pseudo-steady-state approximation. 

Equation (\ref{eq-re-ode}) does not appear to have 
a solution in elementary functions. 
The magnitude of its right-hand side is of order unity, which 
yields an order-of-magnitude estimate $r_e^2(t) \simeq 1 - t/(B_0-1)$. 
Consequently, we have $T \simeq B_0 -1$ and $r_e^2(T/2) \simeq 1/2$. 
Next we obtain a more accurate solution of equation (\ref{eq-re-ode}). 
An approximate polynomial solution would be convenient for comparison 
with the available observational results and theoretical predictions. 
We use a quadratic approximation: 
\begin{equation}
  r_e^2(t) \approx c_0 + 
  c_1 (t-T/2) + c_2 (t-T/2)^2 , 
\label{eq-Taylor}
\end{equation}
where 
\begin{equation}
  c_1 = \left. \frac{d r_e^2}{dt} \right|_{t=T/2}
\end{equation}
and 
\begin{equation}
  c_2 = \frac{1}{2} \left. \frac{d^2 r_e^2}{dt^2} \right|_{t=T/2} . 
\end{equation}
We expand $r_e^2(t)$ about $t=T/2$ because this is where 
the pseudo-steady-state approximation is expected to be most accurate. 
The remaining constants $c_0$ and $T$ are defined by the conditions 
\begin{equation}
  r_e^2 (0) = 1 
\label{eq-r2-t=0}
\end{equation}
and 
\begin{equation}
  r_e^2 (T) = 0 . 
\label{eq-r2-t=T}
\end{equation}
Equations (\ref{eq-Taylor}), 
(\ref{eq-r2-t=0}), and (\ref{eq-r2-t=T}) give 
\begin{equation}
  r_e^2(t) \approx 1 + (c_1 - c_2 T) t + c_2 t^2 . 
\end{equation}
Here 
\begin{equation}
  T = -\frac{1}{c_1} 
\end{equation}
is the sunspot lifetime, unless $T^{\prime} < T$ where 
\begin{equation}
  T^{\prime} = -\frac{c_1}{c_2} 
\end{equation}
is the other root of the equation $r_e^2 (t) = 0$.
 
We evaluate the constant $c_1$ by substituting 
the order-of-magnitude estimates $t = T/2 \simeq (B_0-1)/2$ and 
$r_e^2(T/2) \simeq 1/2$ into equation (\ref{eq-re-ode}). 
This yields an accurate expression for $c_1$ because $T/2$ and 
$r_e^2(T/2)$ only appear in the argument of the logarithm 
in equation (\ref{eq-re-ode}). The resulting prediction 
for the sunspot lifetime is as follows: 
\begin{equation}
  T = \frac{1}{4} (B_0-1) \ln 2(B_0-1) , 
\label{eq-lifetime}
\end{equation}
which should be compared with equation (16) in Petrovay \& Moreno-Insertis (1997) 
for the inward speed $w = - \dot{r}_e$ of the current sheet. In their model, 
$w=\mbox{const}$ and the sunspot lifetime is given by $T_{PM}=r_e/w$, 
which leads to 
\begin{equation}
  T_{PM} = 2^{1/3} B_0 
\label{eq-lifetime-PM}
\end{equation}
in our dimensionless variables. 
The same result (up to a numerical coefficient) is obtained 
by nondimensionalizing our equation (\ref{eq-PM-scaling}), 
substituting it into equation (\ref{eq-stefan}), 
and assuming $\dot{r}_e = \mbox{const}$. 

Differentiation of equation (\ref{eq-re-ode}) 
with respect to time yields 
\begin{equation}
  c_2 = \left. \frac{2}{(B_0-1) [\ln (r_e^2/2t)]^2} 
  \left[ \frac{1}{t} - \frac{4}{(B_0-1) r_e^2 \ln (r_e^2/2t)} 
  \right] \right|_{t=T/2} . 
\end{equation}
Again using $t = T/2 \simeq (B_0-1)/2$ and $r_e^2(T/2) \simeq 1/2$ 
is justified when these quantities appear in the argument 
of the logarithm. Therefore, 
\begin{equation}
  c_2 = \frac{1}{\ln 2(B_0-1)} 
  \left[ 1 + \frac{1}{2 r_e^2(T/2)} 
  \right] \frac{1}{T^2} , 
\label{eq-c2-re2(T/2)}
\end{equation}
where $T$ is defined by equation (\ref{eq-lifetime}). 
The solution below can be used to verify that 
$r_e^2(T/2)=1/2+O(1/\ln 2(B_0-1))$. Thus using 
$r_e^2(T/2) \simeq 1/2$ in equation (\ref{eq-c2-re2(T/2)}) only 
leads to a relatively small error of order $1/[T \ln 2(B_0-1)]^2$, 
and we get 
\begin{equation}
  c_2  \approx \frac{2}{T^2 \ln 2(B_0-1)} . 
\end{equation}
Collecting the results, we obtain a parabolic decay law 
for the sunspot area: 
\begin{equation}
  r_e^2(t) \approx 1 - 
  \left( 1 + \frac{2}{\ln 2(B_0-1)} \right) \frac{t}{T} + 
  \frac{2}{\ln 2(B_0-1)} \frac{t^2}{T^2} . 
\label{eq-our-parabola}
\end{equation}
The sunspot lifetime is given by $T$ in equation (\ref{eq-lifetime}) 
if $B_0 > B_{\ast}$ and by 
\begin{equation}
  T^{\prime} = \frac{1}{8} (B_0-1) \left[ \ln 2(B_0-1) \right]^2 
\label{eq-Tprime}
\end{equation}
if $1 < B_0 < B_{\ast}$, where 
\begin{equation}
  B_{\ast} = 1+e^2/2 \approx 4.7 
\label{eq-Bast}
\end{equation}
corresponds to $T=T^{\prime}$.

Our explicit analytical solution for $r_e (t)$ provides an improved 
quantitative description of sunspot decay by turbulent erosion. 
Notably, if $B_0 = B_{\ast}$, our solution predicts a constant 
decrease rate $w = 2/(B_{\ast}-1) \approx 0.54$ of the fluxtube radius: 
\begin{equation} 
  r_e (t) \approx 1 - wt , 
\end{equation}
as in the parabolic decay law, predicted by Petrovay \& Moreno-Insertis (1997). 
More generally, we obtain a constant speed approximation 
\begin{equation} 
  w \approx 
  \left( \frac{1}{2} + \frac{1}{\ln 2(B_0-1)} \right) \frac{1}{T}
\end{equation} 
by defining $w = -\dot{r}_e(0)$ in our solution. 
If $A(t) = \pi r_e^2$ is the sunspot area, the accuracy of 
the approximation can be quantified by calculating 
\begin{equation}
  \left. \frac{2 \ddot{A}}{\dot{A}^2} \right|_{t=0} = 
  \frac{8 \ln 2(B_0-1)}{[2 + \ln 2(B_0-1)]^2} , 
\end{equation}
which would be unity in the model of Petrovay \& Moreno-Insertis (1997). 

On returning to the original dimensional quantities, 
we get the lifetime--size scaling $T \sim A_0$, where 
$A_0 = \pi r_0^2$ is the initial cross-sectional area of the flux 
tube. This result formally agrees with the Gnevyshev--Waldmeier 
relation for sunspot lifetimes. It is worth stressing that 
the statistical nature of the relation should follow from 
the strong dependence of $T$ on the spot magnetic field $B_0$.

\section{Numerical results}
\label{sec:numerical}

The analytical results obtained in Section~\ref{sec:analytical_model}
may be tested by numerical solution of equation~(\ref{eq-axisymm-diff}).
Following Petrovay \& Moreno-Insertis (1997), we assume 
the analytical forms for the diffusivity and the initial field profile: 
\begin{equation}\label{eq:D(B)}
  D(B)= \frac{1}{1+|B|^{\alpha_D}} , 
\end{equation}
\begin{equation}\label{eq:initial-condition}
  B(r, 0)=\frac{B_0}{1+r^{\alpha_B}} , 
\end{equation}
where we use the non-dimensionalisation introduced in 
Section~\ref{sec:analytical_model}. The parameter $\alpha_D$ in 
equation~(\ref{eq:D(B)}) determines the strength of the suppression of 
diffusion by the field, and the 
parameter $\alpha_B$ in equation~(\ref{eq:initial-condition}) specifies 
the initial spot profile. In the following we choose $\alpha_B=22$
to model an isolated flux tube with nearly constant internal field
strength, and $\alpha_D=7$, to represent strong suppression of 
diffusion. For the purpose of numerical solution, 
the radius of the spot at time $t$ is defined by the condition 
\begin{equation}\label{eq-re-def-num}
B(r_e,t)=\frac{1}{2}B_0.
\end{equation}

We solve equation~(\ref{eq-axisymm-diff}) using a Crank--Nicolson
scheme (e.g.\ Press et al.\ 1992) which is described in the Appendix.
The diffusion equation is evolved in time in the region 
$0\leq r\leq r_m$ with the boundary condition $\partial B / \partial r=0$ 
at $r=0$ and with a boundary condition at $r=r_m$ which allows loss of flux 
from the region. Note that Petrovay \& Moreno-Insertis (1997) used 
a less realistic condition $\partial B / \partial r=0$ 
at an outer boundary (at $r=10$), and their numerical solution 
was based on a Lax--Wendroff scheme. 

\begin{figure*}[h]
\epsscale{0.75}
 \plotone{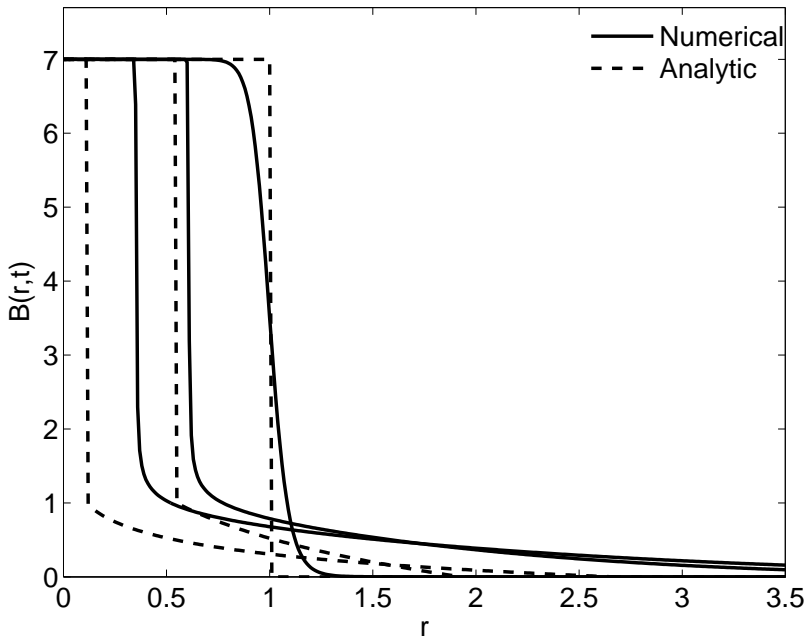}
\caption{Magnetic field versus radius at times $t=0$, $t=0.5T$, and
$t=0.95T$, for the case $B_0=7$. The solid curves show the numerical
solutions (for the parameter choices $\alpha_D=7$, $\alpha_B=22$), and 
the dashed curves show the analytical solutions.}
\label{fig1}
\end{figure*}

Figure~\ref{fig1} illustrates the numerical solution for the case 
$B_0=7$. The solid curves in the figure show the numerical result for
$B(r,t)$ as a function of $r$ for times $t=0$, $t=0.5T$, and 
$t=0.95T$, where $T$ is the analytical decay time, defined
by equation~(\ref{eq-lifetime}).
The solutions are shown for the region $r\leq \frac{1}{2}r_m$,
where $r_m=7$ is the outer boundary of the numerical domain. 
Figure~\ref{fig1} also shows the analytical solution at the same times, 
following equations~(\ref{eq-b-interior}) and~(\ref{eq-b-exterior}) 
with the spot radius defined by equation~(\ref{eq-our-parabola}).
The spot decays more rapidly in the analytical solution, and the magnetic 
field outside the spot decreases more rapidly with increasing radius. 
The numerical solutions illustrate how the initial central flux 
concentration is redistributed to larger radius by diffusion, leading
to an initial increase in field strength at points external to the 
spot. The qualitative behavior of the numerical solution is 
generally well reproduced by the analytical solution.

\begin{figure*}
\epsscale{0.75}
 \plotone{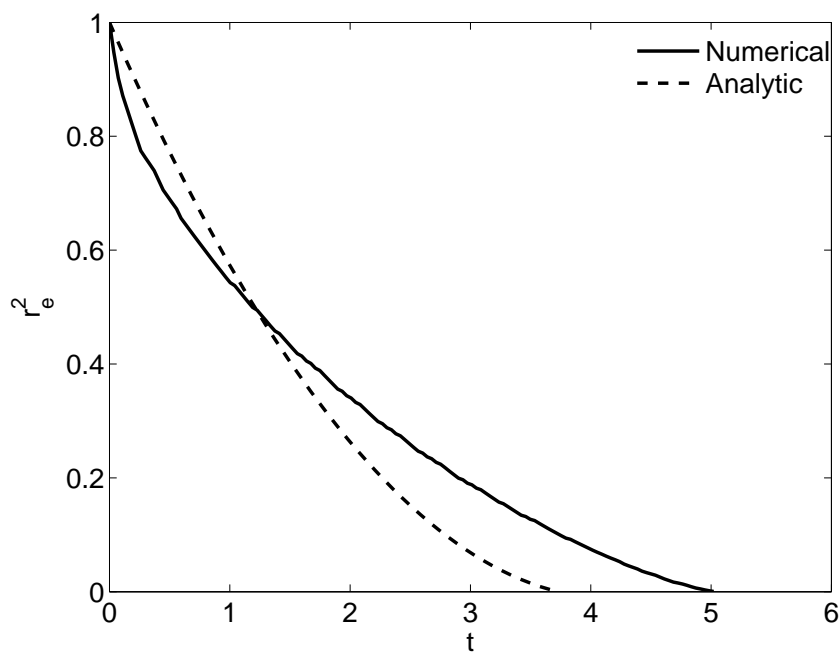}
\caption{Sunspot radius squared versus time, for the case shown in 
Figure~1. The solid curve shows the numerical solution, and the dashed 
curve shows the analytical solution.}
\label{fig2}
\end{figure*}

Figure~\ref{fig2} shows the square of the sunspot radius 
as a function of time for the same case $B_0=7$. 
The solid curve shows the numerical solution, with
$r_e$ defined by equation~(\ref{eq-re-def-num}), and the dashed curve
shows the analytical solution defined by equation~(\ref{eq-our-parabola}).
The analytical solution decays more rapidly than the numerical solution,
but both clearly show the departure from a linear decay law. The 
analytical estimate for the decay time is $T=3.72$, and the numerical
decay time is 5.02. 

\begin{figure*}
\epsscale{0.75}
 \plotone{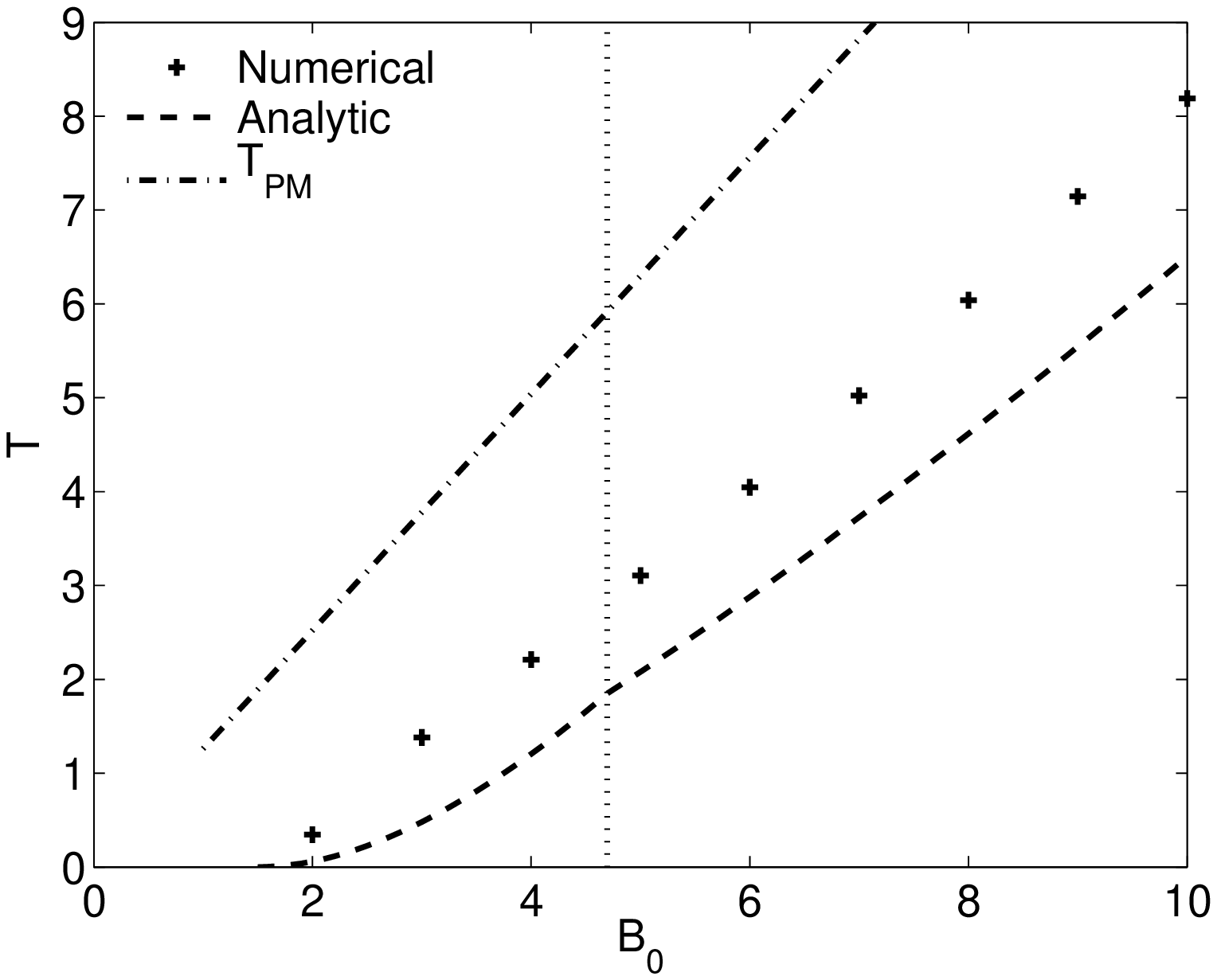}
\caption{Decay time versus sunspot field strength $B_0$. 
The crosses indicate results for numerical solutions 
(with the parameters $\alpha_D=7$, $\alpha_B=22$), 
and the dashed curve is the analytical solution of this paper. 
The dotted vertical line indicates which of two times ($T$ and $T'$) applies. 
The dot-dashed line is the decay time for the Petrovay \& Moreno-Insertis (1997) 
model (our equation~(\ref{eq-lifetime-PM})).}
\label{fig3}
\end{figure*}

Figure~\ref{fig3} plots numerically determined sunspot decay times
versus central field strength $B_0$ (crosses). The decay time is 
seen to depend almost linearly on field strength. The dashed curve 
shows the analytical results of Section~\ref{sec:analytical_model}. 
Recall that equation~(\ref{eq-our-parabola}) defines two times at which 
$r_e (t)=0$, namely $T$ in equation (\ref{eq-lifetime}) 
and $T^{\prime}$ in equation (\ref{eq-Tprime}).
The decay time for the spot is given by $T$ if $B_0 > B_{\ast}$, 
and by $T^{\prime}$ if $B_0 < B_{\ast}$, 
where $B_{\ast}$ is defined by equation (\ref{eq-Bast}). The dotted 
vertical line in Figure~\ref{fig3} indicates the threshold value
$B_{\ast}$. Figure~\ref{fig3} also shows the decay time 
in the Petrovay \& Moreno-Insertis (1997) model. 
Our analytical predictions agree with the numerical results: in particular 
the rates of increase of decay time with $B_0$ are quite similar. 
Although our analytical model underestimates the decay times, 
it is significantly more accurate than the earlier model. 

\begin{figure*}
\epsscale{0.75}
 \plotone{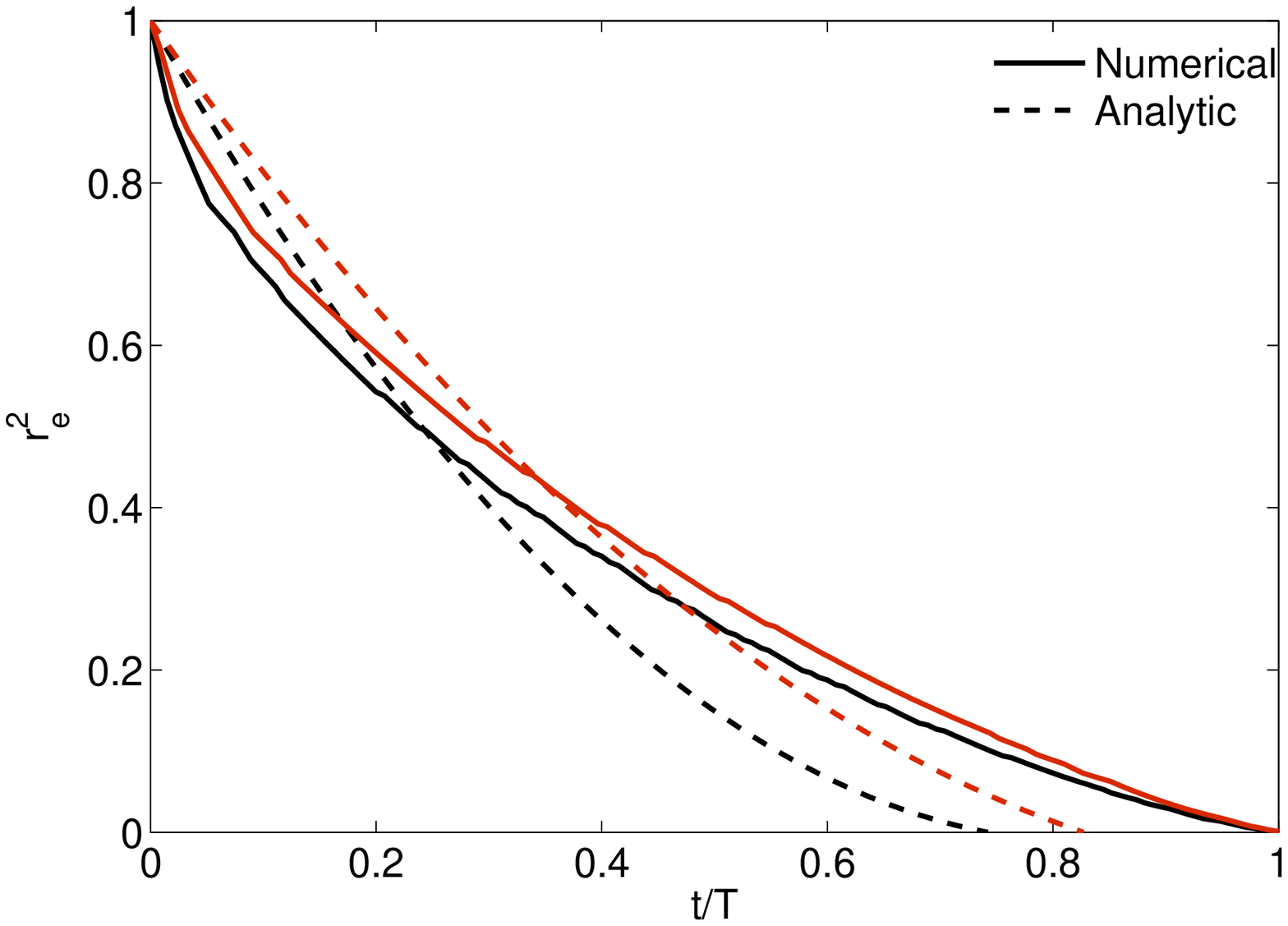}
\caption{Sunspot radius squared versus time, 
normalized by the decay time $T$ from the numerical solution, 
for $ B_0 = 7 $ (black) and $ B_0 = 14 $ (red). 
Other parameters are as in Figure~1. 
The solid curves show the numerical solution, 
and the dashed curves show the analytical solution.}
\label{fig4}
\end{figure*}

Finally, we emphasize that our calculation generally yields 
a time-dependent rate of decrease of the fluxtube radius $r_e(t)$.
Equation (\ref{eq-our-parabola}) predicts that the deviation 
from the parabolic decay law, derived by Petrovay \& Moreno-Insertis (1997), 
should increase as the initial magnetic field $B_0$ increases. 
As a result, a linear decay law (rather than a parabolic one) 
should become more accurate as $B_0$ increases, although the logarithmic dependence 
on $B_0$ makes the effect rather weak. Figure~\ref{fig4} shows the effect 
of doubling the field strength $B_0$ on the shape of the function $r_e^2(t)$. 
While the computation time and numerical errors increase for larger $B_0$, 
we do see numerical evidence that the decay law becomes more linear for 
a larger initial magnetic field, which is consistent with our analytical prediction. 

\section{Discussion}

We have presented in this paper a quantitative theory of sunspot decay 
by turbulent erosion, considered as a moving boundary problem. 
The physical mechanism of sunspot erosion was proposed by 
Simon \& Leighton (1964), and a sunspot decay law due to 
turbulent erosion was derived by Petrovay \& Moreno-Insertis (1997) 
(see also Petrovay et al.\ 1999 and references therein). 
Although Petrovay and collaborators correctly identified the key 
dependence  of the decay rate on the sunspot magnetic field $B_0$, 
the accuracy of the analytical predictions was limited: 
for instance, we have shown that the numerically computed 
sunspot lifetime is about a half of that predicted. 

Our equation (\ref{eq-lifetime}) for the sunspot decay time $T$ 
is an improvement on equation (\ref{eq-lifetime-PM}), 
derived by Petrovay \& Moreno-Insertis (1997). 
Equation (\ref{eq-our-parabola}) confirms that the decay law 
for the sunspot area $A(t)= \pi r_e^2$ is in general parabolic, 
as long as higher-order terms in $t/T$ can be neglected. 
Equation (\ref{eq-our-parabola}) also quantifies the accuracy of the assumption, 
made by Petrovay \& Moreno-Insertis (1997), that the inward speed 
$\dot{r}_e$ of the current sheet surrounding the decaying spot is constant. 
We have shown that the assumption is justified 
if the initial sunspot magnetic field $B_0$ is not too large. 
Equation (\ref{eq-our-parabola}) predicts that a linear decay law 
should become more accurate as $B_0$ increases. 
The numerical solutions in Figure~\ref{fig4} confirm this prediction, 
although, as noted by the referee, they also show that the deviation from 
a linear decay is systematically underestimated in the analytical model.

Application of the turbulent erosion theory to sunspot 
and starspot decay is a topic of current research interest 
(e.g., Strassmeier 2009; Rempel \& Cheung 2014; Bradshaw \& Hartigan 2014), 
and so our quantitative analytical predictions, reinforced by numerical 
solutions, should be useful in studies of solar and stellar activity. 
The value of an analytical calculation is that it can be used to verify 
more detailed magnetohydrodynamic simulations 
(e.g., Hurlburt \& DeRosa 2008; Rempel \& Cheung 2014) 
and to guide empirical models (e.g., Gafeira et al. 2014). 

The erosion model can be further refined. For instance, 
we assumed $D_0 = \mbox{const}$ in our analysis of Section 2. 
The rate of relative diffusion of two photospheric magnetic fragments 
is controlled by turbulent eddies whose size is equal to 
the current distance between the fragments. Consequently, 
the turbulent diffusivity is expected to be scale-dependent. 
In practice the turbulent diffusivity is determined by applying 
the induction equation to pairs of solar magnetograms
(e.g., Chae et al.\ 2008, and references therein). 
Scale-dependent turbulent diffusivity has been invoked to interpret 
observations of photospheric flux cancellation (Litvinenko 2011) and 
the dispersion of photospheric bright points (Abramenko et al.\ 2011). 
The turbulent erosion model of sunspot decay should be 
generalized to incorporate the dependence of the effective 
diffusivity on the size of a decaying sunspot. In addition, 
although Petrovay \& Moreno-Insertis (1997) argued that regular radial flows 
play little if any role in sunspot decay, the effect of regular photospheric 
flows on sunspot decay should be investigated in more detail. 
Finally, recent observations emphasized the difference between 
the maximum and average sunspot magnetic field strengths 
(Tlatov \& Pevtsov 2014), and so it may be worthwhile 
to derive a solution for a more general 
initial profile of the magnetic field within the sunspot, 
as well as a more realistic dependence of the turbulent diffusivity 
on the field strength within the spot. 

\acknowledgments
The authors thank the referee for comments and suggestions that 
helped to improve the original manuscript. 

\appendix

\section{Numerical method}

The numerical solutions in Section~\ref{sec:numerical} use the
Crank--Nicolson method to solve the nonlinear diffusion 
equation~(\ref{eq-axisymm-diff}), in which a discrete version of the
equation is linearised at each time step.
The Crank--Nicolson method is a preferred one for solution of parabolic 
partial differential equations because it is unconditionally stable,
and second order accurate in time (e.g.\ Press et al.\ 1992).

Equation~(\ref{eq-axisymm-diff}) is solved at spatial locations 
$r_j=(j-1)h$ with $j=1,2,\dots,L$ and $h=r_m/(L-1)$, for a sequence 
of times $t_n=(n-1)\tau$, with $n=1,2,\dots$. 
Introducing the notation $B_j^n=B(r_j,t_n)$
and $D_j^n=D(B_j^n)$, we consider a Crank--Nicolson scheme 
with the differencing of terms in equation~(\ref{eq-axisymm-diff}):
\begin{equation}\label{eq-nld-crank-nicolson-working-1}
\left.r\frac{\partial B}{\partial t}\right|_{t_{n},r_{j}}
  \approx r_{j}
  \frac{B_{j}^{n+1}-B_{j}^n}{\tau}
\end{equation}
and
\begin{equation}\label{eq-nld-crank-nicolson-working-2}
\begin{split}
\left.\frac{\partial }{\partial r}
  \left[rD(B)\frac{\partial B}{\partial r}\right]
  \right|_{t_{n},r_{j}}
  &\approx
\frac{1}{2}\left.\frac{\partial }{\partial r}
  \left[rD(B)\frac{\partial B}{\partial r}\right]\right|_{t_{n+1},r_j}
  +\frac{1}{2}\left.\frac{\partial }{\partial r}
  \left[rD(B)\frac{\partial B}{\partial r}\right]\right|_{t_{n},r_j}
  \\
  &\approx 
\frac{1}{2h}\left(
  r_{j+\frac{1}{2}}D_{j+\frac{1}{2}}^{n+1}
    \frac{B_{j+1}^{n+1}-B_j^{n+1}}{h}
 -r_{j-\frac{1}{2}}D_{j-\frac{1}{2}}^{n+1}
    \frac{B_{j}^{n+1}-B_{j-1}^{n+1}}{h}
\right) \\
&\qquad +
\frac{1}{2h}\left(
r_{j+\frac{1}{2}}D_{j+\frac{1}{2}}^{n}
    \frac{B_{j+1}^{n}-B_j^{n}}{h}
 -r_{j-\frac{1}{2}}D_{j-\frac{1}{2}}^{n+1}
    \frac{B_{j}^{n}-B_{j-1}^{n}}{h}
\right).
\end{split}
\end{equation}
In the final expression in 
equation~(\ref{eq-nld-crank-nicolson-working-2}), the centered 
differences are taken about locations $r_{j-\frac{1}{2}}$ and 
$r_{j+\frac{1}{2}}$. We introduce the approximations
$D_{j}^{n+1}\rightarrow D_j^n$ and
\begin{equation}
D_{j\pm\frac{1}{2}}^{n}\rightarrow D_{j\pm}^n=
  \frac{1}{2}\left(D_j^n+D_{j\pm 1}^n\right), 
\end{equation}
involving a linearisation in time and a spatial averaging 
respectively. Combining 
equations~(\ref{eq-nld-crank-nicolson-working-1}) 
and~(\ref{eq-nld-crank-nicolson-working-2}) we have
\begin{equation}\label{eq-cn-update-1}
B_j^{n+1}-\frac{s}{2(j-1)}F(B_j^{n+1})=B_j^n+\frac{s}{2(j-1)}F(B_j^{n}),
\end{equation}
with $s=\tau/h^2$ and
\begin{equation}\label{eq-cn-update-2}
F(B_j^n)=(j-\tfrac{1}{2})D_{j+}^nB_{j+1}^n
  -\left[
   (j-\tfrac{1}{2})D_{j+}^n+(j-\tfrac{3}{2})D_{j-}^n\right]B_j^n
  +(j-\tfrac{3}{2})D_{j-}^nB_{j-1}^n.
\end{equation}
A von Neumann analysis of 
equations~(\ref{eq-cn-update-1})-(\ref{eq-cn-update-2}) in the linear 
case $D_j=D_0=\mbox{const}$ confirms that the scheme is unconditionally 
stable. The corresponding explicit scheme with the same 
spatial differencing is unstable if $D_0\tau/h^2>\frac{1}{2}$
(e.g.\ Press et al.\ 1992).

Equations~(\ref{eq-cn-update-1}) and~(\ref{eq-cn-update-2}) define
the update for points $j=2,3,\dots,L-1$. At the point $j=1$, the
boundary condition $\left.\partial B/\partial r\right|_{r=0} =0$
is enforced using the one-sided second order difference approximation 
to the derivative:
\begin{equation}
\left.\frac{\partial B}{\partial r}\right|_{t_{n+1},r_1}\approx
  \frac{-3B_1^{n+1}+4B_2^{n+1}-B_3^{n+1}}{2h}=0,
\end{equation}
or
\begin{equation}\label{eq-cn-update-3}
-3B_1^{n+1}+4B_2^{n+1}-B_3^{n+1}=0.
\end{equation}

For the point $j=L$ we obtain an update equation allowing flux 
transport across the boundary $r=r_m$ via a discretisation 
of equation~(\ref{eq-axisymm-diff}) at
time $t=t_n$ and spatial location $r=r_{L-\tfrac{1}{2}}$ with
differencing schemes
\begin{equation}\label{eq-cn-update-4-working-1}
\left.r\frac{\partial B}{\partial t}\right|_{t_{n},r_{L-\frac{1}{2}}}
  \approx r_{L-\frac{1}{2}}
  \frac{B_{L-\frac{1}{2}}^{n+1}-B_{L-\frac{1}{2}}^n}{\tau}
\end{equation}
and
\begin{equation}\label{eq-cn-update-4-working-2}
\begin{split} 
  \left.\frac{\partial }{\partial r}
  \left[rD\frac{\partial B}{\partial r}\right]
  \right|_{t_{n},r_{L-\frac{1}{2}}}
&\approx
\frac{1}{h}\left(
   r_LD_L^n\left.\frac{\partial B}{\partial r}\right|_{t_{n},r_{L}}
  -r_{L-1}D_{L-1}^n
    \left.\frac{\partial B}{\partial r}\right|_{t_{n},r_{L-1}}
  \right) \\
&\approx 
  \frac{1}{h}\left(
  r_LD_L^n\frac{B_{L-2}^n-4B_{L-1}^n+3B_L^n}{2h}
 -r_{L-1}D_{L-1}^n\frac{B_{L}^n-B_{L-2}^n}{2h}\right),
\end{split}
\end{equation}
where equation~(\ref{eq-cn-update-4-working-2}) 
involves the one-sided second order difference approximation to the 
derivative:
\begin{equation}
\left.\frac{\partial B}{\partial r}\right|_{t_{n},r_L}\approx
  \frac{B_{L-2}^{n}-4B_{L-1}^{n}+3B_L^{n}}{2h}.
\end{equation}
Equations~(\ref{eq-cn-update-4-working-1}) 
  and~(\ref{eq-cn-update-4-working-2}) 
give the update equation for $j=L$:
\begin{equation}\label{eq-cn-update-4}
\begin{split}
B_L^{n+1}+B_{L-1}^{n+1}&=
  s\left(\frac{L-2}{L-\frac{3}{2}}D_{L-1}^n
  +\frac{L-1}{L-\frac{3}{2}}D_L^n\right)B_{L-2}^n
+\left(1-4s\frac{L-1}{L-\frac{3}{2}}D_L^n\right)B_{L-1}^n\\
  & \qquad\qquad\qquad +\left(1-s\frac{L-2}{L-\frac{3}{2}}D_{L-1}^n
    +3s\frac{L-1}{L-\frac{3}{2}}D_L^n\right)B_L^n . 
\end{split}
\end{equation}

Equations~(\ref{eq-cn-update-1}), (\ref{eq-cn-update-2}), 
(\ref{eq-cn-update-3}), and~(\ref{eq-cn-update-4}) provide a system
of linear equations for the field values $B_j^{n+1}$, with 
$j=1,2,\dots,L$, which must be solved at each time step. 
The scheme may be written in matrix form as
\begin{equation}\label{eq-cn-update-matrix}
\left(\mathrm{I}^{\prime}+a_{-}\mathrm{A}\right){\bf B}^{n+1}
  =\left(\mathrm{I}^{\prime}+a_{+}\mathrm{A}^{\prime}\right)
  {\bf B}^{n},
\end{equation}
where ${\bf B}^n=(B_1^n,B_2^n,\dots,B_L^n)^{T}$,
$a_{\pm}=\mp \frac{1}{2}s$, $\mathrm{I}^{\prime}$ is the $L\times L$ 
matrix
\begin{equation}
\mathrm{I}^{\prime}=\mathrm{diag}(0,1,\dots,1),
\end{equation}
the matrix $\mathrm{A}$ is defined by $A_{L\,j}=0$ for all $j$ except
\begin{equation}
{A}_{L\,L-1}=\frac{1}{a_{-}},
\end{equation}
and the matrix $\mathrm{A}^{\prime}$ is defined by 
$A_{L\,j}^{\prime}=0$ for all $j$ except
\begin{equation}
\begin{split}
A^{\prime}_{L\,L-2}&=\frac{s}{a_{+}}\left(
  \frac{L-2}{L-\frac{3}{2}}D_{L-1}^n+\frac{L-1}{L-\frac{3}{2}}D_L^n
  \right),\\
A^{\prime}_{L\,L-1}&=\left(1-4s\frac{L-1}{L-\frac{3}{2}}D_L^n\right),\\  
A^{\prime}_{L\,L}&=\frac{s}{a_{+}}\left(
  -\frac{L-2}{L-\frac{3}{2}}D_{L-1}^n+3\frac{L-1}{L-\frac{3}{2}}D_L^n
  \right).
\end{split}
\end{equation}

A simple test for the new method is provided by the exact 
solution for constant diffusivity $D_0$ with a Gaussian profile:
\begin{equation}
B(r,t)=\frac{\Phi_0}{\sigma^2}
  \exp\left(-\tfrac{1}{2}r^2/\sigma^2\right)
\end{equation}
with 
\begin{equation}
\sigma^2 = 2D_0t+\sigma_0^2 ,
\end{equation}
where the constant $\sigma_0$ defines the initial width. 
The magnetic flux (divided by $2 \pi$) from $r=0$ to $r=r_m$ 
for this solution is
\begin{equation}
\begin{split}\label{eq-gauss-sol-flux}
\Phi (r_m,t)&=\int_0^{r_m}rB(r,t)\,dr\\
  &=\Phi_0\left[1-\exp\left(-\tfrac{1}{2}r_m^2/\sigma^2\right)\right].
\end{split}
\end{equation}
Equation~(\ref{eq-gauss-sol-flux}) provides a check on the
implementation of the boundary condition at $r=r_m$. The method incurs
truncation error (proportional to $\tau^2$ and $h^2$) at each time step,
and the accumulation of the error limits the accuracy of the solution 
when the system is evolved over many time steps. The calculations 
presented in this paper are checked by trials with different spatial steps.

%\begin{thebibliography}{}
\begin{flushleft}
%\bibitem{} 
Abramenko, V. I., Carbone, V., Yurchyshyn, V., et al. 2011, 
  ApJ, 743, 133 
\\
%\bibitem{}
Bradshaw, S. J., \& Hartigan, P. 2014, 
  ApJ, 795, 79
\\
%\bibitem{}
Bumba, V. 1963, 
  Bull. Astron. Inst. Czech., 14, 91
\\
%\bibitem{}
Carslaw, H. S., \& Jaeger, J. C. 1959, 
  Conduction of Heat in Solids (Oxford: Clarendon Press)
\\
%\bibitem{}
Chae, J., Litvinenko, Y. E., \& Sakurai, T. 2008, 
  ApJ, 683, 1153
\\
%\bibitem{}
Chatterjee, P., Choudhuri, A. R., \& Petrovay, K. 2006, 
  A\&A, 449, 781
\\
%\bibitem{}
Crank, J. 1984, 
  Free and Moving Boundary Problems (Oxford: Clarendon Press)
\\
%\bibitem{}
Gafeira, R., Fonte, C. C., Pais, M. A., \& Fernandes, J. 2014, 
  Sol. Phys., 289, 1531
\\
%\bibitem{}
Hill, J. M., \& Dewynne, J. N. 1987, 
  Heat Conduction (Oxford: Blackwell Scientific)
\\
%\bibitem{}
Hurlburt, N., \& DeRosa, M. 2008, 
  ApJL, 684, L123
\\
%\bibitem{}
Kitchatinov, L. L., Pipin, V. V., \& R\"udiger, G. 1994, 
  Astron. Nachr., 315, 157
\\
%\bibitem{}
Krause, F., \& R\"udiger, G. 1975, 
  Sol. Phys., 42, 107
\\
%\bibitem{}
Litvinenko, Y. E. 2011, 
  ApJL, 731, L39
\\
%\bibitem{}
Meyer, F., Schmidt, H. U., Weiss, N. O., \& Wilson, P. R. 1974, 
  MNRAS, 169, 35
\\
%\bibitem{}
Mart\'inez Pillet, V., Moreno-Insertis, F., \& V\'azquez, M. 1993, 
  A\&A, 274, 521
\\
%\bibitem{}
Moreno-Insertis, F., \& V\'azquez, M. 1988, 
  A\&A, 205, 289
\\
%\bibitem{}
Petrovay, K., Mart\'inez Pillet, V., \& van Driel-Gesztelyi, L. 1999, 
  Sol. Phys., 188, 315
\\
%\bibitem{}
Petrovay, K., \& van Driel-Gesztelyi, L. 1997, 
  Sol. Phys., 176, 249
\\
%\bibitem{}
Petrovay, K., \& Moreno-Insertis, F. 1997, 
  ApJ, 485, 398
\\
%\bibitem{}
Press, W. H., Flannery, B. P., Teukolsky, S. A., \& Vetterling, W. T.
  1992, Numerical Recipes in C: The Art of Scientific Computing,
  Second Ed.\ (Cambridge: Cambridge Univ.\ Press)
\\
%\bibitem{}
Rempel, M., \& Cheung, M. C. M. 2014, 
  ApJ, 785, 90
\\
%\bibitem{}
Robinson, R. D., \& Boice, D. C. 1982, 
  Sol. Phys., 81, 25 
\\
%\bibitem{}
R\"udiger, G., \& Kitchatinov, L. L. 2000, 
  Astron. Nachr., 321, 75
\\
%\bibitem{}
Simon, G. W., \& Leighton, R. B. 1964, 
  ApJ, 140, 1120
\\
%\bibitem{}
Solanki, S. K. 2003, 
  Astron. Astrophys. Rev., 11, 153
\\
%\bibitem{}
Strassmeier, K. G. 2009, 
  Astron. Astrophys. Rev., 17, 251 
\\
%\bibitem{}
Tlatov, A. G., \& Pevtsov, A. A. 2014, 
  Sol. Phys., 289, 1143

\end{flushleft}
%\end{thebibliography}
\end{document}